\documentstyle[prl,aps,multicol,epsfig]{revtex}

\def\kf{k_{\mbox{\tiny{F}}}}              

\def\Ef{E_{\mbox{\tiny{F}}}}

\def\vec#1{{\bf #1}}

\begin{document}

\title{Oscillating Sign of Drag in High Landau Levels}

\author{Felix von Oppen$^1$, Steven H. Simon$^2$, and Ady Stern$^3$}

\address{$^1$ Institut f\"ur Theoretische Physik, Freie Universit\"at
Berlin, Arnimallee 14, 14195 Berlin, Germany \\ $^2$ Lucent
Technologies, Bell Labs, Murray Hill, NJ, 07974\\ $^3$ Department of 
Condensed Matter Physics, The Weizmann Institute of  Science, 
76100 Rehovot, Israel}

\date{April 30th, 2001}

\maketitle

\begin{abstract} 
  
  Motivated by experiments, we study the sign of the Coulomb drag
  voltage in a double layer system in a strong magnetic field. We show
  that the commonly used Fermi Golden Rule approach implicitly assumes
  a linear dependence of intra-layer conductivity on density, and is
  thus inadequate in strong magnetic fields. Going beyond this
  approach, we show that the drag voltage commonly changes sign
  with density difference between the layers. We predict that in the
  Quantum Hall regime the Hall and longitudinal drag resistivities are
  comparable. Our results are also relevant for pumping and
  acoustoelectric experiments.

\end{abstract}      




\begin{multicols}{2}  

Drag experiments in coupled two-dimensional electron systems provide
information on the response of a system at finite frequency and
wavevector and are thus complementary to standard DC transport
measurements \cite{MacDonaldZheng,DragTheory}. In a typical drag
experiment, a current is applied to the {\it active} layer of a
double-layer system and the voltage $V_D$ induced in the other {\it
passive} layer is measured, with no current allowed to flow in
that layer.  In a simple picture of drag, the current in the active
layer leads -- via {\it inter-layer} Coulomb or phonon interaction
-- to a net transfer of momentum to the carriers in the passive
layer. At conditions of zero current in the passive layer, this
momentum transfer is counteracted by the build-up of the drag
voltage $V_D$. In the cases of two electron layers or two hole
layers, the drag voltage points opposite to the voltage drop in the
active layer. This is defined as positive drag. Negative drag occurs
in systems with one electron layer and one hole
layer\cite{DragTheory,electronhole}.

Prior theoretical work on drag \cite{DragTheory,DragHighB,PhononDrag}
was often based on or reduced to a Fermi Golden Rule analysis (see,
e.g., Zheng and MacDonald \cite{MacDonaldZheng}). In this analysis,
the sign of drag does not vary with magnetic field $B$, temperature
$T$, or difference in Landau level (LL) filling factor $\nu$ between
the two layers.  By contrast, recent experiments at large
perpendicular $B$ \cite{Gramila,Lok,Sivan} observe that the sign of
drag changes with all of these parameters in systems of two coupled
electron layers -- specifically in the Shubnikov-de-Haas (SdH) and
integer quantum Hall (IQH) regimes. Feng {\it et al.}  \cite{Gramila}
find positive drag whenever the topmost partially filled LLs in both
layers are either less than half filled or more than half filled.
Negative drag is observed when the topmost LL is less than half-filled
in one layer and more than half-filled in the other. Even more
surprisingly, Lok {\it et al.} \cite{Lok} maintained that the sign of
drag was sensitive to the relative orientation of the majority spins
of the two layers at the Fermi energy.

In this paper, we show that the use of the Fermi Golden Rule
approach is inappropriate in the SdH and IQH regimes and that a more
careful analysis opens different routes to drag with changing
sign. Remarkably, we find that Hall drag can be of the same magnitude
as longitudinal drag in these regimes. We emphasize that a naive
rationale for the experimental results of Feng {\it et al.} -- which
points to the similarity between a less (more) than half-filled Landau
level and an electron (hole) like band -- leaves out essential physics
of the problem. Finally, we compare our results to the existing
measurements \cite{Gramila,Lok} and propose some interesting
experiments.

We start from a general linear-response expression \cite{DragTheory}
which relates the drag conductivity $\hat\sigma^D$ to the
rectification coefficients $\bf\Gamma({\bf q},\omega)$ of the active
and passive layers,
\begin{eqnarray}
        \sigma^D_{ij}=\!\int \frac{d \omega}{2\pi}\sum_{\vec q} 
         &&\frac{|U(\vec q,\omega)|^2}{8T \sinh^2(\omega/2 T)}
    \nonumber\\
        &&\hspace*{.5cm}\times\Gamma^p_i(\vec q,\omega;B) 
           \Gamma_j^a(\vec q,\omega;-B).
        \label{eq:sigmaD}
\end{eqnarray}
The drag conductivity gives the current response ${\bf J}^p$ in the
passive (dragged) layer via $J^p_i = \sigma^D_{ij} E^a_j$ to an
applied electric field ${\bf E}^a$ in the active layer. Its relation
to the commonly measured drag resistivity is clarified below. In Eq.\
(\ref{eq:sigmaD}), $U(\vec q,\omega)$ is the screened interlayer
(Coulomb or phonon-mediated) interaction and the rectification
coefficient ${\bf \Gamma}(\vec q,\omega)$ is defined by
\begin{equation}        
\label{eq:Jdc}        
\vec J^{dc} = \frac{1}{2}\sum_{\vec q, \omega} \,\vec \Gamma(\vec q,\omega) 
|e\phi(\vec q,\omega)|^2.
\end{equation} 
In Eq. (\ref{eq:Jdc}), $\vec J^{dc}$ is the DC current induced in
quadratic response to the driving force exerted by a screened potential
$\phi(\vec q, \omega)$ of wave-vector ${\bf q}$ and frequency
$\omega$.

The physical interpretation of the drag expression Eq.\ 
(\ref{eq:sigmaD}) is that the voltage in the active layer creates an
asymmetry (between $\vec q$ and $-\vec q$) in the thermal density
fluctuations in that layer.  These fluctuations are transferred to the
passive layer -- via Coulomb or phonon interaction --- where they are
rectified to create a current. In experiment, usually the drag
resisitivity $\hat\rho^D$ is measured, which gives the electric field
response $\vec E^p$ in the passive layer to a current $\vec J^a$
driven in the active layer via $E^p_i = \rho^D_{ij} J^a_j$. For weak
drag, it is related to the drag conductivity by
\begin{equation}
        \label{eq:rhoD}
        \hat \rho^D = \hat \rho^p \hat \sigma^D 
        \hat \rho^a
\end{equation} 
where $\hat \rho^{p(a)}$ is the resistivity tensor of the passive 
(active) layer. 

A simple but instructive approach to rectification considers
situations in which current and field are locally related, $\vec
J(\vec r,t)=-\hat \sigma(n(\vec r,t))\nabla \phi(\vec r,t)$, and the
conductivity depends on position and time only through the local
density $n({\bf r},t)$ \onlinecite{Esslinger}.  In addition to the
current, the perturbation $\phi(\vec q,\omega)$ also induces a density
perturbation $\delta n(\vec q,\omega) = \Pi(\vec q,\omega) e\phi(\vec
q,\omega)$ due to the polarizability $\Pi({\bf q},\omega)$.  Up to
quadratic order in the applied potential,
\begin{eqnarray}
 \vec J(\vec r,t) = -\left[\hat \sigma(n_0) +
\frac{d \hat \sigma}{d n}\delta n(\vec r,t)  \right]
 \nabla \phi(\vec r,t).
\end{eqnarray}
Taking the time and space average of the second term then yields a DC
contribution to the current given by
\begin{equation}
\vec J_{dc} =  \sum_{{\bf q},\omega}\left(\frac{d
\hat \sigma}{dn} \right) \left[\Pi(\vec q,\omega)e \phi(\vec
q,\omega)
i \vec q  \phi(-\vec q, -\omega)\right],
\end{equation}
or equivalently the rectification coefficient is given by
\begin{equation}
        \label{eq:mainresult}
  {\bf \Gamma}(\vec q,\omega) = -2
    {d\hat\sigma\over d(e n)}\cdot {\vec q}\,
   {\rm Im}\Pi(\vec q,\omega).
\end{equation}  
At zero magnetic field, the conductivity is local when considering
scales large compared to the mean free path $\ell_{\rm el}$, implying
that Eq.\ (\ref{eq:mainresult}) holds in the diffusive regime defined
by $q\ell_{\rm el}\ll 1$ \cite{coherence}. At finite $B$, and for
short range disorder, Eq.\ (\ref{eq:mainresult}) is still valid in the
diffusive regime $\omega\tau,Dq^2\tau\ll1$, where $D$ and $\tau$
denote the appropriate diffusion constant and scattering time.
Indeed, we have checked this \cite{tobepublished} by an explicit
diagrammatic calculation in the self-consistent Born approximation
(SCBA) \cite{Ando}. Specifically, once the magnetic field is strong
enough such that the cyclotron radius $R_c= \hbar \kf/(e B)$ is
small compared to $\ell_{\rm el}$ and for short-range correlated
disorder, the conductivity is local on scales larger than $R_c$, and
Eq.\ (\ref{eq:mainresult}) holds for $qR_c\ll 1$ \cite{coherence}.
Since $R_c\ll \ell_{\rm el}$ even for very small magnetic fields, and
since the diffusion constant decreases rapidly with the application of
a magnetic field, Eq.\ (\ref{eq:mainresult}) is expected to be
applicable for a significant part of the wave-vector range of the
integration in (\ref{eq:sigmaD}) for typical experiments.

Using Eq.\ (\ref{eq:mainresult}) in the general drag expressions, Eqs.\
(\ref{eq:sigmaD}) and (\ref{eq:rhoD}), one readily finds that up to an
overall {\it positive} prefactor, the drag resistivity tensor becomes
\begin{equation}
   \hat\rho^D \sim \hat\rho^p{d \hat\sigma^p\over d(en)}
       {d \hat\sigma^a\over d(en)} \hat\rho^a.
\label{eq:drag-resistivity}
\end{equation}
This expression easily reproduces some standard results.  In the
absence of a magnetic field, the tensor structure is trivial.
Observing that the conductivity increases (decreases) with increasing
electron density for electron (hole) layers, we recover that coupled
layers with the same type of charge carriers exhibit positive drag,
while coupled electron-hole layers have a negative drag resistivity.
If the system is strictly electron-hole symmetric, the derivative of
the conductivity with respect to density vanishes and consequently,
there is no drag.  Treating the conductivity in the presence of a
magnetic field in a simple Drude-type picture, $\hat\sigma$ depends
linearly on the density $en$ (so long as the scattering time is taken
to be density independent). Eq.\ (\ref{eq:mainresult}) then reproduces
results of previous works\cite{MacDonaldZheng,DragTheory,PhononDrag}
that calculate drag in a Fermi Golden Rule or scattering time
approximation.  In particular, in this approximation it is immediately
found from Eq.\ (\ref{eq:drag-resistivity}) that the drag resistivity
is diagonal and Hall drag vanishes identically. Finite Hall drag can
appear even in a Drude-type approximation once the scattering time is
taken to be energy and thus density dependent \cite{Halldrag}.

Interesting new effects appear in the presence of a strong magnetic
field, in the SdH and IQH regimes, where the derivative of $\sigma_{xx}$
changes sign as the magnetic field or Fermi energy is varied.  While
this is superficially quite reminiscent of the experimental results of
Refs.\ \onlinecite{Gramila,Lok}, the details can be involved due to
the many terms that contribute, once the tensor products in Eq.\
(\ref{eq:drag-resistivity}) are multiplied out.  In the experimental
samples, typically $\rho_{xy}\gg\rho_{xx}$ already at very small
magnetic fields. Thus, up to an overall positive prefactor
\begin{equation}
  \rho^D_{xx}\sim \rho^p_{xy}\left\{{d\sigma^p_{yy}\over d(en)}
    {d\sigma^a_{yy}\over d(en)}+{d\sigma^p_{yx}\over d(en)}
    {d\sigma^a_{xy}\over d(en)}\right\}\rho_{yx}^a.
\label{eq:drag-exp}
\end{equation}
Generally the derivative of the longitudinal conductivity
$\sigma_{xx}$ changes sign in the SdH and IQH regimes, being positive
for less than half filling of the topmost LL and negative for more
than half filling. By contrast, the Hall conductivity generally
increases monotonically with $n$, and its derivative is therefore
positive.

It is not clear, apriori, which of the two terms in the curly brackets
of Eq.\ (\ref{eq:drag-exp}) dominates.  We obtain an oscillatory sign
of drag, similar to the experimental results in Ref.\ \cite{Gramila}
when $d\sigma_{xx}/d(en)$ dominates over $d\sigma_{xy}/d(en)$.
However, if that were the case, drag would be {\it negative for equal
filling of the two layers} (because $\rho_{xy}=-\rho_{yx}$), in
contrast to the experimental observations.  Positive drag for equal
filling is only obtained when $d\sigma_{xy}/d(en)\gg
d\sigma_{xx}/d(en)$. Were that the case, however, there would
presumably be no sign changes of the drag resistivity.  In fact, in
the IQH regime the derivatives of both components of the conductivity
tensor are experimentally of the same order (both change by
approximately $e^2/h$ in the region of the plateau transition). While
neither of the two limits is therefore realized in experiment, it is
clear that Eq.\ (\ref{eq:drag-exp}) makes it difficult to obtain both
positive drag for equal densities and a drag sign that oscillates
with the difference in densities.

A striking consequence of Eq.\ (\ref{eq:drag-resistivity}) is that if
we assume that $d\sigma_{xx}/d(en)$ and $d\sigma_{xy}/d(en)$ are of
comparable magnitude in the integer-quantum-Hall regime then the Hall
drag is of the same order as longitudinal drag. In fact, we have
\begin{equation}
  \rho^D_{xy}\sim \rho^p_{xy}\left\{{d\sigma^p_{yy}\over d(en)}
    {d\sigma^a_{yx}\over d(en)}+{d\sigma^p_{yx}\over d(en)}
    {d\sigma^a_{xx}\over d(en)}\right\}\rho_{xy}^a.
\label{eq:Hall-drag-exp}
\end{equation}
Moreover, Hall drag generally changes sign with filling factor
difference between the two layers.

Although the experimental disorder potential is long ranged, it is
instructive to consider also the case of short-range disorder, for
which the SCBA \cite{Ando} becomes {\it exact\ }in the limit of high
LLs \cite{Chalker}. Assuming that the disorder broadening is small
compared to the LL spacing, the conductivity tensor $\hat\sigma$
becomes \cite{Ando} $\sigma_{xx}=(e^2/\pi^2\hbar)
N\{1-[(E_F-E_N)/2\gamma]^2\}$ and $\sigma_{xy}=-(en/B)+ (e^2/\pi^2
\hbar)(2\gamma/\hbar\omega_c)N\{1-[(E_F-E_N)/2\gamma]^2\}^{3/2}$ where
$\gamma$ denotes the LL broadening, and the Fermi energy $E_F$ is
assumed to lie within the $N$-th LL (of energy $E_N = (N+1/2) \hbar
\omega_c$ in the clean case).  Thus
$|d\sigma_{xx}/d(en)|\gg|d\sigma_{xy}/d(en)|$ and $d\sigma_{xx}/d(en)$
changes sign as a function of filling -- being positive for less than
half filled LLs and negative for more than half-filled LLs. Neglecting
the derivatives of the Hall conductivity, one obtains for the diagonal
drag resistivity of an isotropic system
\begin{equation}
\label{eq:res1}
   \rho^D_{xx}=  C_{xx}(B,T)
   \frac{d\sigma^p_{xx}}{d(en)}\frac{d\sigma^a_{xx}}{d(en)} [
   \rho^p_{xx}\rho^a_{xx} -    \rho^p_{xy}\rho^a_{xy}]
\end{equation}
with $C_{xx}$ a positive function. Thus, we again find the surprising
result that drag is {\it negative} for equal densities in the two
layers (since $|\rho_{xy}/\rho_{xx}|\ge\pi$). More generally, drag
oscillates with difference in density between the layers. It is
negative whenever the topmost occupied LLs in the two layers are both
less than half filled or both more than half-filled. It is positive if
the topmost LL is less than half-filled in one layer and more than
half-filled in the other.  Moreover, the Hall drag resistivity is
comparable to the diagonal drag resistivity also in this model
situation. In fact, one has
\begin{equation}
\label{eq:res2}
  \rho^D_{xy} = C_{xy}(B,T)
      \frac{d\sigma^p_{xx}}{d(en)}
\frac{d\sigma^a_{xx}}{d(en)}[\rho^p_{xy}\rho^a_{xx}
      +\rho^p_{xx}\rho^a_{xy}],
\end{equation}
with $C_{xy} = C_{xx}$. Unlike before, this result is now a
consequence of the fact that $\rho_{xx}$ and $\rho_{xy}$ are of
similar magnitudes.

When the LL broadening is comparable to but exceeds the LL spacing
($\omega_c\tau<1$), we find \cite{tobepublished} from a numerical
evaluation of the SCBA equations that by contrast, the derivative of
$\sigma_{xy}$ dominates over that of $\sigma_{xx}$ while still
changing sign as function of filling (for
$N\gamma/\hbar\omega_c\gg1$).  However, in this limit, $\sigma_{xx}$
dominates over $\sigma_{xy}$ (and hence $\rho_{xx}$ over $\rho_{xy}$)
so that again, we find negative drag for equal filling of both
layers. 

Experiments do not necessarily satisfy the condition $Dq^2\tau\ll
1$ for the diffusive regime. Thus, we now turn to a discussion of drag
in the ballistic regime where $Dq^2\tau\gg 1$.  At zero $B$, Eq.\ 
(\ref{eq:mainresult}) holds also in the ballistic regime
\cite{DragTheory}.  Indeed, Eq.\ (\ref{eq:mainresult}) with $\hat
\sigma$ given by the Drude expression is also derivable within a
Boltzmann approach which should be valid for $q/\kf , \omega/\Ef \ll
1$ and at fields low enough that SdH oscillations are absent
\cite{tobepublished}.  Surprisingly, we find that this is not
generally true for higher magnetic fields.  While we have not
succeeded in deriving a general expression, analogous to Eq.\ 
(\ref{eq:mainresult}), in this regime, we computed the rectification
explicitly in the high-magnetic field limit introduced above.

Our calculation starts from the diagrammatic expression for the
rectification coefficient\cite{DragTheory}
\begin{eqnarray}
  {\bf\Gamma}({\bf q},\omega)={\omega\over\pi i}
    {\rm Tr}\left\{{\bf I}{ G}^-e^{i{\bf qr}}[{G}^--{G}^+]e^{-i{\bf qr}}
    {G}^+\right\}
\end{eqnarray}
Here, $G^\pm$ denotes the impurity-averaged Green function in the
SCBA, ${\bf I}$ is the current operator, and ${\rm Tr}$ the trace over
the single-particle states.  The calculation is simplified by the fact
that vertex corrections can be neglected in the ballistic limit.

Using standard results for the matrix elements of the current and
density operators between LL wavefunctions and exploiting the small
parameter $\gamma/\hbar\omega_c\ll 1$, we obtain for the longitudinal
rectification
\begin{eqnarray}
\label{eq:parballistic}
   \Gamma_\parallel = \left( {-4 \omega R_c\over N}\right) J_0(qR_c)
    J_1(q R_c) \left({dn\over d\mu}\right)^2 {d\sigma_{xx}\over d(en)}.
\end{eqnarray} 
with $J_i$ the $i^{th}$ Bessel function and $N$ the LL number. A similar 
calculation for the transverse rectification gives $\Gamma_\perp=(1/\omega_c\tau)
\Gamma_\parallel$ or
\begin{eqnarray}
\label{eq:perpballistic}
   \Gamma_\perp=\frac{2}{3} \left( \frac{-4 \omega R_c}{N}\right) J_0(qR_c) 
    J_1(q R_c) \left({dn\over d\mu}\right)^2  {d\sigma_{xy}\over d(en)} .
\end{eqnarray}
While both $\Gamma_\parallel$ and $\Gamma_\perp$ still include factors
that can be written as derivatives of the corresponding conductivity,
similar to the diffusive regime, the prefactors are no longer equal to
one another and cannot be expressed in terms of the polarization
operator $\Pi(q,\omega)$. Remarkably, the rectification in the
ballistic limit also changes sign with $qR_c$ due to the matrix
elements of the density, in addition to the sign changes of
$d\hat\sigma/d(en)$ discussed above.  Note that such behavior would be
impossible in Eq.\ (\ref{eq:mainresult}) since the sign of ${\rm Im}
\Pi$, being fixed by the sign of $\omega$, does not change with $q$.

It is an interesting question whether these additional sign changes
are reflected in the drag. This is not obvious since the oscillating
Bessel functions depend on $q$ which needs to be integrated over to
obtain the drag conductivity (cf.\ Eq.\ (\ref{eq:sigmaD})).  For
Coulomb drag, the screened interlayer interaction can be approximated
by the Thomas-Fermi result $U({\bf q},\omega)=\pi e^2
q/(\kappa_a\kappa_p\sinh qd)$ with $\kappa_{a}$ ($\kappa_p$) denoting
the Thomas-Fermi momenta of the active (passive) layer and $d$ the
interlayer distance.  Thus, the $\omega$ and $q$ integrations in Eq.\ 
(\ref{eq:sigmaD}) factorize. One readily observes that the $q$
integration is {\it not} dominated by the upper limit in the relevant
region $1/R_c\ll q\ll 1/d$, and thus there are no additional
oscillations from the matrix elements in Coulomb drag. This result is
also confirmed by straight-forward numerical evaluation.

For phonon drag, the situation is slightly different. Using a
qualitative approximation \cite{PhononDrag},
$|U(q,\omega)|^2 \approx K q^3 \delta(\omega - q v)$ with $v$
the sound velocity, the integrals in Eq.\ (\ref{eq:sigmaD}) 
yield an additional factor $T^4 [g(q_T |R_c^a - R_c^p|) + g(q_T 
|R_c^a + R_c^p|)]$ where $q_T = (T/T_B) \kf$ with $T_B = 2 \hbar v \kf$
the Bloch-Gr\"uneisen temperature (for typical samples $T_B \approx 10$
K).  Here, $g(x)$ is a smooth function (which can be calculated analytically) 
that starts with $g(0) = 1$ and decreases 
for small $x$ crossing zero at $x \approx .04$ then reaching a minimum of $g(.075) \approx -.37$ then increasing monotonically and exponentially towards zero.   For
all but the lowest temperatures, the term involoving $|R_c^a - R_c^p|$  
dominates, and thus the overall sign
of the drag changes when $q_T |R_c^a - R_c^p| = |\nu_a -
\nu_p| T/T_B \approx .04$.

In passing, we note that while the sign changes of ${\bf \Gamma}({\bf
  q},\omega)$ with $qR_c$ in the ballistic limit seem to have only
minor consequences for drag, they should be observable in other
experiments of recent interest, namely in pumping \cite{Pump} and in
acoustoelectric experiments \cite{Esslinger}, which also measure the
rectification coefficient ${\bf \Gamma}({\bf q},\omega)$
\cite{Aleiner}.  In pumping experiments, a potential $\phi(\vec r,t)$
is applied to the system which is typically made up of the sum of two
potentials oscillating out of phase $\phi(\vec r,t) = \phi_1(\vec r)
\cos(\omega t) + \phi_2(\vec r) \cos(\omega t + \delta)$.  Due to the
symmetries of ${\bf \Gamma}({\bf q},\omega)$, the pumping current
density is given by $\vec J^{dc} = (1/2)\sin(\delta) \sum_{\vec q}
\vec \Gamma(\vec q,\omega) {\rm Im}[\phi_1(\vec q) \phi_2(-\vec q)]$.
In acoustoelectric experiments, acoustic waves sent through a
(piezoelectric) crystal apply an ``external'' electric potential
$\phi^{ext}(\vec q,\omega)$ to the electrons in the system with
$\omega = c q$ and $c$ the wave velocity.  The driven electric current
is given precisely by Eq.\ (\ref{eq:Jdc}) where $\phi({\bf q},\omega)
=\phi^{ext}(q,\omega)/[1-v(q)\Pi(q,\omega)]$ is the screened potential
associated with $\phi^{ext}$ ($v(q)$ denotes the bare Coulomb
interaction).  We note that the acoustoelectric experiments of Ref.\ 
\onlinecite{Esslinger}, which were carried out in the diffusive
regime, have been successfully analysed using Eq.\ 
(\ref{eq:mainresult}).

For the cases considered here, an oscillating sign of drag is always
accompanied by negative drag for equal fillings. This apparent
conflict with the experimental results of Feng {\it et al.} might be
resolved in one of the following ways: (a) For long-range disorder and
$\hbar\omega_c$ comparable to the LL broadening, the Hall conductivity
might become non-monotonous, cf.\ Eq.\ (\ref{eq:drag-resistivity}) and
the SCBA results.  (b) In measurements, the large Hall drag
resistivity might mix into the measured drag response.  On the other
hand, a simple-minded extension of our results to include the two spin
directions definitely fails to account for the apparent spin
dependence observed in Ref.\ \cite{Lok}.  We take this failure as an
indication that the observations of Ref.\ \cite{Lok} cannot be
explained by theories where correlations induced by Coulomb
interactions are neglected.

In conclusion, we have considered drag in high LLs with an emphasis on
the sign of the effect. Remarkably, we find that both in the ballistic
and the diffusive regimes, drag in high LLs {\it cannot} be described
by a widely-used Fermi Golden Rule expression \cite{MacDonaldZheng}.
Our analysis naturally opens the possibility of sign changes of the
drag resistivity as function of filling factor difference between the
two layers.  Moreover, it implies that Hall drag can be of the same
order as longitudinal drag in high LLs.  Surprisingly, we find that
the sign of drag can be quite different from naive expectations.  In
particular, in several regimes drag is {\it negative} for two
identical electron layers. We believe that it would be particularly
interesting to check experimentally our predictions that Hall drag can
be of the same order as longitudinal drag and that the acoustoelectric
current in the ballistic regime changes sign as function of the
wavevector.

We thank B.I.\ Halperin, K.\ von Klitzing, S.\ Lok, M.\ Raikh, U.\ 
Sivan, and Y.\ Yaish for instructive discussions. We acknowledge
financial support from the US-Israel BSF, the Israeli science
foundation, DIP-BMBF (AS), and through the DFG-Schwerpunkt
``Quanten-Hall-Systeme'' (FvO).  FvO thanks the Einstein Center and
the LSF at the Weizmann Institute for hospitality.

\end{multicols}


\begin{references}
\bibitem{coherence} Here we neglect phase coherence which leads to
small nonlocal contributions on scales of the coherence length.
\end{references}
\end{document}